\documentclass[aps,prl,twocolumn,showpacs]{revtex4-1}
\usepackage{amsmath}
\usepackage{graphicx,epsfig}
\usepackage{bm}

\renewcommand{\v}[1]{{\bf #1}}
\newcommand{\be}{\begin{equation}}
\newcommand{\ee}{\end{equation}}
\newcommand{\bd}{\begin{displaymath}}
\newcommand{\ed}{\end{displaymath}}
\newcommand{\ba}{\begin{eqnarray}}
\newcommand{\ea}{\end{eqnarray}}
\newcommand{\nn}{\nonumber \\}

\newcommand{\bpm}{\begin{pmatrix}}
\newcommand{\epm}{\end{pmatrix}}

\newcommand{\ie}{{\it i.e.}}

\begin{document}

\title{Spin-orbit Coupled Bose-Einstein Condensate under Rotation}

\author{Xiao-Qiang Xu and Jung Hoon Han}
\email[Electronic address:$~~$]{hanjh@skku.edu}
\affiliation{Department of Physics and BK21 Physics Research Division,
Sungkyunkwan University, Suwon 440-746, Korea}

\date{\today}

\begin{abstract}
We examine the combined effects of Rashba spin-orbit (SO) coupling and
rotation on trapped spinor Bose-Einstein condensates (BECs).
Nature of single particle states is thoroughly examined in
the Landau level basis and is shown to support the formation of half-quantum vortex.
In the presence of weak $s$-wave interactions,
the ground state at strong SO coupling develops ring-like
structures with domains whose number shows step behavior with
increasing rotation. For fast rotation case, the
vortex pattern favors triangular lattice, accompanied by the
density depletion
in the central region and weakened Skyrmionic character as
the SO coupling is enhanced. Giant vortex formation is facilitated
when SO coupling and rotation are both strong.
\end{abstract}
\maketitle

\textit{Introduction}.- Bose-Einstein condensates (BECs) are a
fascinating testing ground for many interesting condensed matter
physics phenomena such as the formation of vortices by rotation. For
a single component condensate, if the rotation is fast enough, the
vortices form a triangular (Abrikosov) lattice, indicating the entry
of BECs into a quantum Hall regime where the lowest Landau level
(LLL) physics dominates~\cite{vortex}. In the case of spinor
condensates, the competition between inter- and intra-component
interactions was found to yield even richer vortex patterns
such as square lattice, double-core lattice, and interwoven
``serpentine" vortex sheet~\cite{two-component-VL}, or even giant Skyrmions~\cite{two-component-Mason} if
the symmetry between the two components is broken.

A recent focus in cold atom physics is on the influence of
``engineered" Rashba-type or Dresselhaus-type spin-orbit (SO) coupling among spinor
BECs induced by the so-called ``synthetic non-Abelian gauge
fields"~\cite{GaugeField}.
The SO effects in cold atoms have received increasing attention
theoretically~\cite{galitski,zhai,spin-2} and its first experimental
realization was recently reported~\cite{rashba}. Non-trivial
new structures, such as stripe phase and half-quantum vortex, have
been found or predicted, enriching the phase diagram of spinor BEC
system~\cite{zhai,WuAndZhai}.

All existing studies of SO effect in spinor condensates refer to
non-rotating case, and the aim of this paper is to study the
combined effects of Rashba SO coupling and rotation on spinor BECs,
focusing on spin-1/2 case as the prototype. Even the single particle
aspect of the Rashba BEC Hamiltonian in the presence of rotation is
not well understood theoretically, and we address these issues
first. We use the Landau level (LL) expansion to examine the ground state of the single particle
system by exact diagonalization, then move on
to discuss the influences of weak $s$-wave collisional interactions.
New features which are absent from pure Rashba, or pure rotational
BECs, are observed.
\\

\textit{Single Particle States}.- We start with the single particle
Hamiltonian describing the trapped Rashba SO coupled system under rotation
$H_0 = \int \Psi^\dag (\v p^2 + 2\kappa \v p \cdot \bm \sigma) \Psi/(2m) -  \Omega_z L_z + \int \Psi^\dag V(\v r) \Psi$,
where $\v p = -i \hbar \bm \nabla$ with $m$ being the particle mass, $\bm \sigma$ are the $2\times2$ Pauli matrices,
and $\kappa$ denotes the strength of SO coupling.
Note that we make implicit the $2\times2$ unit matrix for $\v p$.
$\Psi = (\psi_\uparrow, \psi_\downarrow)^T$
describes the two-component wave function. $\Omega_z$ stands for the rotational frequency and $L_z =
\int \Psi^\dag (\v r \times \v p) _z \Psi$ is the canonical angular momentum, while the non-canonical
part $\propto \kappa \v r \times \bm \sigma/m$ can be compensated for an external spatially dependent Rabi coupling
between the two components~\cite{MechanicalLz-Wu}.
The trapping potential $V(\v r) = m \omega^2 (x^2+y^2)/2$ is
introduced to keep the boundary condition realistic. We assume that
the trapping frequency in the $z$ direction is much higher than those of
transverse directions, so we may view our condensate as an effective
two-dimensional system with the degree of freedom in the $z$
direction frozen out. With the proliferation of parameters in the
problem, it is convenient to turn all quantities dimensionless by
introducing a length scale $\sqrt{\hbar/m \omega}$ and an energy
scale $\hbar \omega$.  The dimensionless Hamiltonian thus becomes
\ba H_0\!\! &=& \!\! {1\over 2} \!\int | (-i \bm
\nabla + \gamma \bm \sigma - b \hat{z} \times \v r ) \Psi |^2 \nn
&& + \gamma   b  \! \int \! \Psi^\dag ( \hat{z} \times \v r \cdot \bm
\sigma ) \Psi + {1\over 2} ( 1 - b^2 ) \!\int\! r^2 \Psi^\dag \Psi.
\label{eq:H-rotating-Rashba} \ea
Two dimensionless parameters $\gamma = \kappa/\sqrt{\hbar m \omega}$
and $b=\Omega_z /\omega$ are introduced. Note that the second term of Eq.~(\ref{eq:H-rotating-Rashba})
is induced by the interplay between rotation and SO effects. We require $b<1$ in order
to prevent the complete cancelation of the trapping potential by the
centrifugal force.

\begin{figure}[ht]
\includegraphics[width=88mm]{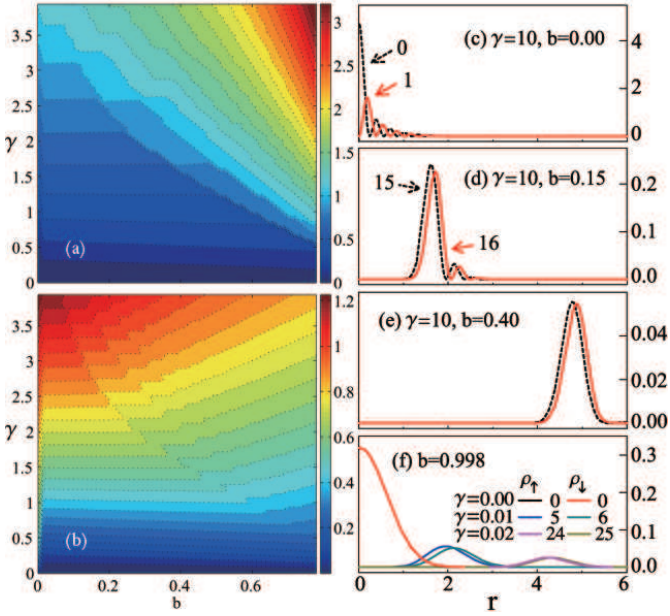}
\caption{(Color online) Single particle ground state of rotating
Rashba-BEC Hamiltonian. (a) Average angular momentum ${\langle
\hat{c}^\dag \hat{c} \rangle}$ and (b) average LL index ${\langle
\hat{a}^\dag \hat{a} \rangle}$ for $\psi_\uparrow$
in $(b,\,\gamma)$ parameter space. Cubic roots ($\langle
\hat{c}^\dag \hat{c} \rangle^{1/3}$ and $\langle \hat{a}^\dag
\hat{a} \rangle^{1/3}$) are plotted for improved view. (c) to (e)
are radial density distributions of $\psi_\uparrow$ (black dotted) and
$\psi_\downarrow$ (red solid) for the same large $\gamma$ and increasing $b$. The numbers
with arrows denote the winding numbers of wave
functions. (f) Densities of both components for the
same $b$ and increasing $\gamma$. The digits in the legend
label the winding numbers. } \label{fig:singleParticle}
\end{figure}

The solution of the single particle Hamiltonian~(\ref{eq:H-rotating-Rashba}) is
conveniently obtained from its second-quantized form which reads
\ba \hat{H}_0 \!=\! (1\!+\!b)\hat{a}^\dag \hat{a} \!+\!
(1\!-\!b)\hat{c}^\dag \hat{c}+i \gamma \bpm 0 & \hat{a}^\dag \!-\! \hat{c} \\
\hat{c}^\dag \!-\! \hat{a} & 0 \epm, \label{eq:2nd-quantized-H} \ea
where operators are introduced as
$\hat{a}={1\over 2} z +\partial_{\bar{z}}$,
$\hat{a}^\dag= {1\over 2} \bar{z}-\partial_{z}$,
$\hat{c}=  {1\over 2} \bar{z} +  \partial_{z}$, and
$\hat{c}^\dag={1\over 2} z  - \partial_{\bar{z}}$
in complex coordinates $z=x+iy$, $\bar{z}=x-iy$, $\partial_z = (\partial_x - i
\partial_y )/2$, and $\partial_{\bar{z}} = (\partial_x + i
\partial_y )/2$. The two pairs of operators satisfy the bosonic commutation
relations, $[\hat{a},\hat{a}^\dag ]=[\hat{c},\hat{c}^\dag ]=1$.
We focus on positive $b$ and $\gamma$ without losing generality.

Without SO coupling, each Landau level state would be an eigenstate
satisfying $\hat{a}^\dag \hat{a} \phi_{n,m}=n \phi_{n,m}$ and $
\hat{c}^\dag \hat{c} \phi_{n,m}=m \phi_{n,m}$ with $n$ and $m$ being the Landau level index
and angular momentum number. The eigenenergy is $\varepsilon_0 = (1+b)n+(1-b)m$, $\ie$, $(1+b)$ and
$(1-b)$ are the energy spacing between LLs and angular momentum states in each LL, respectively.
The ground state is achieved for $n=m=0$. For nonzero $\gamma$, on the other hand,
due to the specific
form of Rashba term which couples neighboring LLs, one can use the ansatz that the eigenstate
be formed as the linear combination $\Psi = \sum_{m \ge
\nu} (x_m \phi_{m-\nu+1, m},\,y_m \phi_{m-\nu, m})^T$. By substituting $\Psi$ into
$\hat{H}_0 \Psi = \varepsilon \Psi$  with
 $\varepsilon$ being the eigenenergy, we find that
the coefficients $x_m, y_m$ obey
\ba
&&[(1\!+\!b)(m\!\!-\!\!\nu\!+\!1) + (1\!-\!b) m ] x_m \nn
&& ~~~+ i\gamma ( \sqrt{m\!\!-\!\!\nu\!+\!1} y_m -
\sqrt{m\!+\!1} y_{m\!+\!1} )= \varepsilon x_m , \nn
&&[(1\!+\!b)(m\!-\!\nu) + (1\!-\!b) m ] y_m \nn
&& ~~~+ i\gamma ( \sqrt{m} x_{m\!-\!1} - \sqrt{m\!\!-\!\!\nu
\!+\!1} x_{m} )= \varepsilon y_m . \ea
The one-dimensional coupled linear equation set can be solved by
numerical exact diagonalization with a reasonably large cutoff in
$m$ for each $\nu$. Note that $\nu$ is an integer number defining
the phase windings of $\psi_\uparrow$ and $\psi_\downarrow$ as $\nu-1$ and $\nu$, respectively, as shown
in Fig.~\ref{fig:singleParticle}(c)-\ref{fig:singleParticle}(f). The difference by 1 indicates the formation of
half-quantum vortex~\cite{HQV}.
 When either
$b$ or $\gamma$ increases, the value of $\nu$ in the ground state increases correspondingly.
In the limit $b = 1$, each LL becomes infinitely degenerate, and we find by numerical approach the divergence of $\nu$  due to
the disappearance of trapping potential.

In Fig.~\ref{fig:singleParticle}(a)-\ref{fig:singleParticle}(b)
we show the average angular momentum $\langle \hat{c}^\dag \hat{c}
\rangle$ and average LL index $\langle \hat{a}^\dag \hat{a}
\rangle$ for $\psi_\uparrow$ ($\psi_\downarrow$
shows similar behavior). The visible steps in figures are strongly
related to the discreteness of LLs. For fixed $\gamma$,
increasing $b$ results in higher angular momentum and lower LL index due
to the decrease of energy spacing between angular momentum states within each LL $(1-b)$ and the increase of
energy spacing between LLs $(1+b)$.
It also results in the more
ring-like structure of the ground state density with increasing
radius at larger $b$. By calculating the local spin density vector
$\v n = \Psi^\dag \bm \sigma \Psi $, $\ie$,
$n_x = \psi^*_\uparrow \psi_\downarrow + \psi^*_\downarrow \psi_\uparrow$,
$n_y = -i \psi^*_\uparrow \psi_\downarrow + i \psi^*_\downarrow \psi_\uparrow$,
$n_z = \psi^*_\uparrow \psi_\uparrow - \psi^*_\downarrow \psi_\downarrow$,
we also find that the spin
texture at $b=0$ corresponds to the Skyrmion
configuration which can be characterized by the Skyrmion density
\ba\label{eq:skyrDen}
\rho_s (\v r) = \frac{1}{4\pi}\frac{ \v n \cdot (\partial_x \v n \times \partial_y \v n )}{|\v n|^3}.
\ea
As $b$ increases,
we observe the growing overlap between densities of two components, resulting in the
decrease of $\rho_s$ due to $n_z \rightarrow 0$. Note that the Skyrmion density is strictly zero
for any planar spin texture. Along with the density depletion in the trap center, the ground state
exhibits spin spirals in the annular region.

When SO coupling is absent, $b\rightarrow 1^-$ limit is associated with
the quantum Hall regime. For
small $\gamma$, we may view the SO coupling as a perturbation
and the LLL picture still applies. However, when $\gamma$ increases to be
comparable to $(1-b)$, higher angular momentum
states contribute
more and more to the ground state and the ring-like nature as well
as a large winding number $\nu$ becomes dominant as seen
in Fig.~\ref{fig:singleParticle}(f). For the value of $b$ we choose, when $\gamma$
is comparable to the LL energy spacing $(1+b)$, the radius becomes enormously large.

\textit{Interaction Effects}.- We may now include interactions in
the manner suitable for spin-1/2 condensate,
\ba H_\mathrm{I} = {1\over 4} \int  [g_c (\rho_\uparrow +
\rho_\downarrow )^2 - g_s (\rho_\uparrow - \rho_\downarrow )^2 ],
\label{eq:H-interaction} \ea
where $g_c = (g_{\uparrow\uparrow} + g_{\uparrow\downarrow})N$ and
$g_s = (g_{\uparrow\downarrow} - g_{\uparrow\uparrow})N$ with
$g_{\uparrow\uparrow}$ and $g_{\uparrow\downarrow}$ being the intra-
and inter-component interactions, respectively. It is assumed that
$g_{\uparrow\uparrow}=g_{\downarrow\downarrow}$ and
$g_{\uparrow\downarrow}=g_{\downarrow\uparrow}$. $N$ is the particle
number, thus $\rho_i=|\psi_i|^2$ ($i=\uparrow,\,\downarrow$)
satisfy  $\int \sum_i \rho_i = 1$. The full Hamiltonian becomes $H =
H_0 + H_\mathrm{I}$. The ground state is numerically
obtained using the imaginary time evolution method. Now we also
consider two specific cases as in the single particle analysis
above.

\begin{figure}[ht]
\includegraphics[width=85mm]{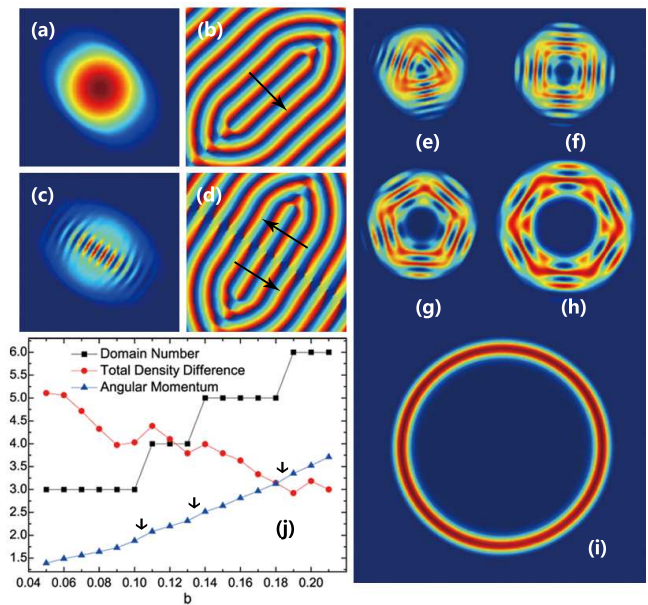}
\caption{(Color online) Ground state evolution with $b$ when
$\gamma=10$ and $(g_c, g_s ) = (9, -1)$.  (a) and (b) are plots of the
density $|\psi_\uparrow|^2$ and phase $\mathrm{arg}(\psi_\uparrow)$ for $b=0.0$,
while (c) and (d) are for $b=0.02$. The long arrows in (b) and (d) represent the directions of $\v k$
vectors. (e)-(i): Up-component densities $|\psi_\uparrow|^2$
for $b=$ 0.08, 0.12, 0.16, 0.20, and 0.50, respectively. (j) Dependence of domain number, total
density difference $\int |n_z|$ ($\times 10$), and angular momentum of
$\psi_\uparrow$ on $b$. $\psi_\downarrow$ shares
similar evolution behavior.}\label{fig:pw}
\end{figure}

We start with strong SO coupling case. When there is no
rotation, the ground state is a stripe (ST) ($g_s > 0$) or a plane
wave (PW) ($g_s < 0$) as was found recently~\cite{zhai}. In our trapped system, we
choose weak interactions so as to recover the same stripe and PW phases.
 We first
consider $g_s < 0$ case and discuss $g_s > 0$ later. The PW phase is
composed of a single $\v k$ state as shown in Fig.~\ref{fig:pw}(a)-\ref{fig:pw}(b).
As $b$ increases gradually above zero, the $-\v k$
state also begins to contribute, and each component of the condensate is divided
into two parts with opposite momenta, indicated by long arrows in Fig.~\ref{fig:pw}(d). Vortices
emerge due to the counter-propagation of plane waves. Locations of these vortices also coincide with
the low density regions shown in Fig.~\ref{fig:pw}(c). We notice the
similarity of this phase to the metastable state found in
Ref.~\cite{zhai}. As we keep on increasing $b$, more $\v k$ states
contribute, and a single domain of vortex line
(Fig.~\ref{fig:pw}(c)) turns into several symmetrically placed
domains with different $\v k$'s (Fig.~\ref{fig:pw}(e)-\ref{fig:pw}(h)), while
vortices nucleate at the center of the trap. The number of different
$\v k$ domains increases as $ 3\rightarrow 4\rightarrow 5\rightarrow
6 \rightarrow \cdots$ until each component of the condensate becomes
indistinguishable from a concentric ring (giant vortex) at strong enough rotation
as shown in Fig.~\ref{fig:pw}(i).
The ring then expands in radius as
$b$ further increases. The integer number of
domains shows step behavior with rotation as shown in
Fig.~\ref{fig:pw}(j). Additionally, the angular momentum of the
condensate increases monotonously with $b$, showing slight jumps
(indicated by short arrows in Fig.~\ref{fig:pw}(j)) where the domain
number changes. We also observe the enhanced overlap between
densities of two components as $b$ increases. We define $\int
|n_z|$ as the
total density difference to quantify the overlap, where the absolute value is chosen
to avoid cancelation among opposite signs of $n_z$ upon the integration. From Fig.~\ref{fig:pw}(j)
we observe the obvious overall trend of decrease, where the minor turns also coincide with
the transition points of the domain number.

\begin{figure}[ht]
\includegraphics[width=88mm]{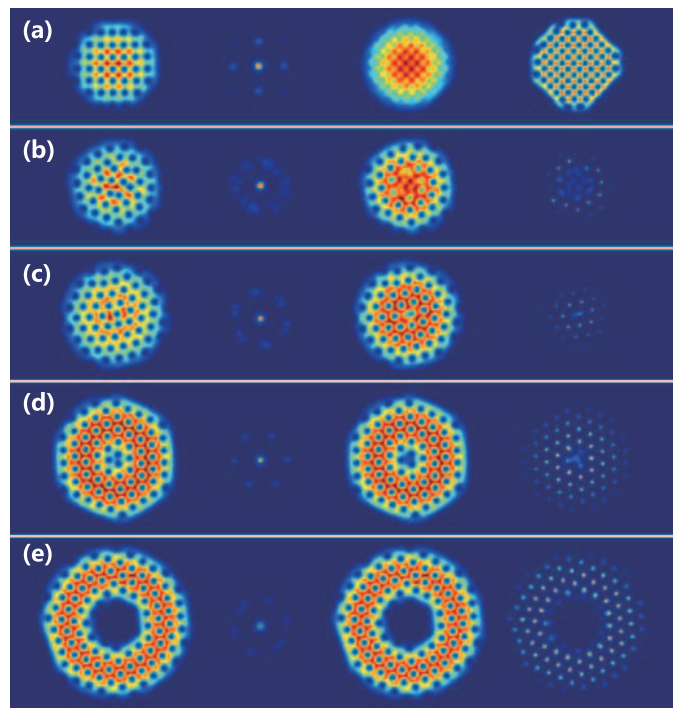}
\caption{(Color online) Ground state evolution with $\gamma$ for large rotation $b=0.998$ and $(g_c,\,g_s ) = (9,-1)$.
From left to right: density of up-component $|\psi_\uparrow|^2$,
magnitude of its Fourier transformation $|\int \mathrm{exp}[-i \v k\cdot \v r]|\psi_\uparrow|^2|$, total density of two components
$(|\psi_\uparrow|^2 + |\psi_\downarrow|^2)$, and the
Skyrmion density $\rho_s$. (a)-(e):
$\gamma$=0.0, 0.002, 0.1, 0.2, and 0.3, while the total Skyrmion number,
defined as the spatial integral of Eq.~(\ref{eq:skyrDen}), is 40.9143,
19.2982, 12.8454, 8.1136, and 3.7456, respectively. }\label{fig:square}
\end{figure}

So far, we have discussed $g_s<0$ case. For $g_s>0$, the ground state is a
stripe phase corresponding to the standing wave formed by two
counter-propagating plane waves when $b=0$. As we increase rotation,
we also observe similar evolution behavior as already shown in
Fig.~\ref{fig:pw}(c)-\ref{fig:pw}(i). This implies that for weak interactions,
as $b$ increases, the difference caused by the sign of $g_s$ is
smeared out, giving way to the effects of large $\gamma$ and $b$.

When there is no SO coupling,  the ground state in the so-called
quantum Hall regime ($b\rightarrow 1^-$) is largely composed of the
LLLs. Previous theories for the symmetrical case $(g_{\uparrow\uparrow}=g_{\downarrow\downarrow})$ found an interlaced vortex
array for each component, while the detailed lattice
structures depend on the
value of $g_s$~\cite{two-component-VL}. Starting from the known
limit, $\gamma=0$, we hereby examine the influence of finite
$\gamma$. For consistency, we choose the same values of
interactions as in the strong SO coupling case and discuss the influence
of $g_s$ later. Square vortex lattice
($g_s \rightarrow 0^-$) is shown in
Fig.~\ref{fig:square}(a), along with the corresponding Skyrmion density
$\rho_s (\v r)$. Interestingly, $\gamma =0$
lattice of displaced vortex cores of $\psi_\uparrow$ and $\psi_\downarrow$ already
exhibits a regular pattern of Skyrmion density, qualifying it as a
Skyrmion lattice~\cite{cornell}.
We emphasize that the same Skyrmion charge is obtained for both spin-up and spin-down vortex cores.

From top to bottom of Fig.~\ref{fig:square} we can find that, with
increasing $\gamma$, the vortex lattice would be distorted first,
and then densities deplete away from the trap center, forming
the ring-like structure as in the single particle case. In the
annular region of Fig.~\ref{fig:square}(d)-\ref{fig:square}(e) the vortices prefer to
form triangular lattice pattern, even though we started with the
square lattice for $\gamma=0$. Inside the trap the giant vortex is formed.
During the transition from square to triangular lattice, we observe that the Skyrmion
distribution shrinks first, and then expands as $b$ increases as shown in Fig.~\ref{fig:square}. To characterize the
trend of the Skymrion density becoming more squeezed, we define
the total Skyrmion number as $\int \rho_s$ which decreases correspondingly
(see the caption of Fig.~\ref{fig:square}). This fact results from the
increased overlap between up- and down-component densities. When the
vortices overlap completely, Skyrmions are also destroyed.

Ground state of pure rotating spin-1/2 BECs may also be interlaced triangular vortex lattice when
$g_s$ is close to $-g_c$
($g_{\uparrow\downarrow}$ is small compared with $g_{\uparrow\uparrow}$),
or double-core lattice when $g_s=0$,
or even vortex sheet when $g_s>0$~\cite{two-component-VL}. For weak interactions, as we increase SO
coupling, we observe similar evolution pattern
as shown in Fig.~\ref{fig:square} for all these vortex structures. Ring structure
with a giant vortex about the trap center would appear, while triangular vortex lattices
is favored in the annulus, and the overlap between densities would be enhanced. As in the case of strong SO coupling
discussed above, the details of $g_s$ play no qualitative role for large $\gamma$ and $b$.
\\

\textit{Summary}.- We
examined several prominent features of the rotating trapped Rashba BECs by
analytical and numerical means.  As we increase the rotational frequency, the
system favors lower LLs with higher angular momenta, while the increase of SO coupling strength would drive the
system to higher LLs with higher angular momentum. The two trends result from the competition between
 SO coupling strength and the energy spacings in the LL system. We also found the winding number difference by 1 between
the two components indicating the formation of half-quantum vortex.

Weak interactions modify the single
particle ground state dramatically. At strong SO coupling, the ring-structured densities develop multiple domains
whose number shows step behavior as we increase the rotational
frequency. At strong rotation, increasing SO
coupling strength would favor triangular vortex lattice,
deplete densities from the trap center, and weaken the Skyrmion
crystal order. In the trap center the giant vortex is formed when SO coupling and rotation are both strong. Very recently
in Ref.~\cite{Realization} we noticed the discussion of possible schemes
for rotating SO coupled BEC system, exhibiting the challenge of experimental realization. Our findings
are expected to build the understanding of the interplay between
SO coupling and rotation.

\acknowledgments H. J. H. is supported by Mid-career Researcher
Program through NRF grant funded by the MEST (No.
R01-2008-000-20586-0). We acknowledge helpful discussions and many
insightful comments on our draft from Hui Zhai.

\end{document}